\documentclass[12pt]{article}

\usepackage{fullpage,amsfonts,amsmath,amsthm,amssymb,mathrsfs,graphicx}
\usepackage[usenames,dvipsnames]{color}

\usepackage[margin=1in]{geometry}

\begin{document}



\def\dd{r}
\def\ld{{\widehat L}}
\def\Gcal{\mathcal G}
\def\eps{\epsilon}
\def\eqn{\eqref}
\def\wtau{\widehat \tau}
\def\ttau{\widetilde \tau}
\def\Q{\mathcal Q}
\def\hG{\widehat H}
\def\C{\mathcal C}
\def\ds{{\mathcal D}}
\def\br{br}
\def\M{{\cal M}}
\def\H{{\cal H}}
\def\bell{\ensuremath{\boldsymbol\ell}}
\def\Bin{{\bf Bin}}


\newcommand{\be}{\begin{equation}}
\newcommand{\ee}{\end{equation}}
\newcommand{\bea}{\begin{eqnarray}}
\newcommand{\eea}{\end{eqnarray}}
\newcommand{\bean}{\begin{eqnarray*}}
\newcommand{\eean}{\end{eqnarray*}}
\newcommand{\non}{\nonumber}
\newcommand{\no}{\noindent}
\newcommand\floor[1]{{\lfloor #1 \rfloor}}
\newcommand\ceil[1]{{\lceil #1 \rceil}}
\newcommand{\remove}[1]{}
\newcommand{\lab}[1]{\label{#1}\ }

\def\a{\alpha}
\def\b{\beta}
\def\c{\chi}
\def\d{\delta}
\def\D{\Delta}
\def\e{\epsilon}
\def\f{\phi}
\def\F{\Phi}
\def\g{\gamma}
\def\G{\Gamma}
\def\k{\kappa}
\def\K{\Kappa}
\def\z{\zeta}
\def\th{\theta}
\def\Th{\Theta}
\def\l{\lambda}
\def\la{\lambda}
\def\La{\Lambda}
\def\m{\mu}
\def\n{\nu}
\def\p{\pi}
\def\P{\Pi}
\def\r{\rho}
\def\R{\Rho}
\def\s{\sigma}
\def\S{\Sigma}
\def\t{\tau}
\def\om{\omega}
\def\Om{\Omega}
\def\smallo{{\rm o}}
\def\bigo{{\rm O}}
\def\to{\rightarrow}
\def\E{{\bf Exp}}
\def\ex{{\mathbb E}}
\def\cd{{\cal D}}
\def\rme{{\rm e}}
\def\hf{{1\over2}}
\def\R{{\bf  R}}
\def\cala{{\cal A}}
\def\cale{{\cal E}}
\def\call{{\cal L}}
\def\cald{{\cal D}}
\def\calz{{\cal Z}}
\def\calf{{\cal F}}
\def\Fscr{{\cal F}}
\def\cc{{\cal C}}
\def\calc{{\cal C}}
\def\calh{{\cal H}}
\def\calk{{\cal K}}
\def\cals{{\cal S}}
\def\calr{{\cal R}}
\def\calt{{\cal T}}
\def\msq{{\mathscr Q}}
\def\bk{\backslash}

\def\out{{\rm Out}}
\def\temp{{\rm Temp}}
\def\overused{{\rm Overused}}
\def\big{{\rm Big}}
\def\moderate{{\rm Moderate}}
\def\swappable{{\rm Swappable}}
\def\candidate{{\rm Candidate}}
\def\bad{{\rm Bad}}
\def\crit{{\rm Crit}}
\def\col{{\rm Col}}
\def\dist{{\rm dist}}
\def\poly{{\rm poly}}

\def\tdT{{\widetilde\Theta}}

\newcommand{\Exp}{\mbox{\bf Exp}}
\newcommand{\var}{\mbox{\bf Var}}
\newcommand{\pr}{\mbox{\bf Pr}}

\newtheorem{lemma}{Lemma}
\newtheorem{theorem}[lemma]{Theorem}
\newtheorem{conjecture}[lemma]{Conjecture}
\newtheorem{corollary}[lemma]{Corollary}
\newtheorem{claim}[lemma]{Claim}
\newtheorem{remark}[lemma]{Remark}
\newtheorem{proposition}[lemma]{Proposition}
\newtheorem{observation}[lemma]{Observation}
\newtheorem{property}[lemma]{Property}
\theoremstyle{definition}
\newtheorem{definition}[lemma]{Definition}

\newcommand{\limninf}{\lim_{n \rightarrow \infty}}
\newcommand{\proofstart}{{\bf Proof\hspace{2em}}}
\newcommand{\tset}{\mbox{$\cal T$}}
\newcommand{\proofend}{\hspace*{\fill}\mbox{$\Box$}}

\newcommand{\bfm}[1]{\mbox{\boldmath $#1$}}
\newcommand{\reals}{\mbox{\bfm{R}}}
\newcommand{\expect}{\mbox{\bf Exp}}
\newcommand{\he}{\hat{\e}}
\newcommand{\card}[1]{\mbox{$|#1|$}}
\newcommand{\rup}[1]{\mbox{$\lceil{ #1}\rceil$}}
\newcommand{\rdn}[1]{\mbox{$\lfloor{ #1}\rfloor$}}
\newcommand{\ov}[1]{\mbox{$\overline{ #1}$}}
\newcommand{\inv}[1]{\frac{1}{#1} }
\newcommand{\imax}{I_{\rm max}}

\newcommand{\whp}{w.h.p.}
\newcommand{\aas}{a.a.s.}

\date{\empty}

\title{Inside the clustering window for random linear equations}

\author{Pu Gao\footnote{Major part of this research was done when the author was affiliated with University of Toronto, supported by an NSERC Postdoctoral Fellowship.} \\
School of Mathematical Sciences\\
Monash University\\
jane.gao@monash.edu
\and Michael Molloy\footnote{Research supported by an NSERC Discovery Grant and Accelerator Supplement.} \\
Department of Computer Science\\
University of Toronto\\
molloy@cs.toronto.edu}

\maketitle

\begin{abstract}  We study a random system of  $cn$ linear equations over $n$ variables in GF(2), where each equation contains exactly $r$ variables; this is equivalent to $r$-XORSAT. Previous work has established a clustering threshold, $c^*_r$ for this model: if $c=c_r^*-\e$ for any constant $\e>0$ then with high probability all solutions form a well-connected cluster; whereas if $c=c^*_r+\e$, then {with high probability} the solutions partition into well-connected, well-separated {\em clusters}  (with probability tending to 1 as $n\rightarrow\infty$).  This is part of a general clustering phenomenon which  is hypothesized to arise in  most of the commonly studied models of random constraint satisfaction problems, via sophisticated but mostly non-rigorous techniques from statistical physics.
We extend that study to the range $c=c^*_r+o(1)$, and prove that the connectivity parameters of the $r$-XORSAT clusters undergo a smooth transition around the clustering threshold.

\end{abstract}

\section{Introduction}
\lab{sec:intro}

 The study of random constraint satisfaction problems (CSP's) has been revolutionized by a collection of hypotheses, arising from statistical physics, concerning the geometry of solutions. According to these hypotheses, before reaching the satisfiability threshold of a random CSP (e.g.\ $r$-SAT), the geometry of its solution space undergoes several phase transitions. {Roughly speaking: there are specific constants, including the {\em clustering threshold}, the {\em freezing threshold}, and the {\em condensation threshold}, all below the satisfiability threshold (see~\cite{mmbook} for an overview). These phase transitions indicate dramatic changes in the degree of the correlation between solutions as the density of a random CSP grows; which shed insights into why it is challenging to determine the satisfiability threshold for many CSP's, or to find efficient CSP solvers when the density is close to the satisfiability threshold. This paper focuses on the clustering threshold. When the density (the ratio of the number of constraints to $n$, the number of variables) of a random CSP instance exceeds the clustering threshold, the set of solutions \aas\footnote{ A property holds \aas\ (asymptotically almost surely)
if it holds with probability tending to 1 as $n\rightarrow\infty$.} partitions into exponentially many clusters, whereas before the density reaches the clustering threshold, all solutions form a single cluster.  One can move throughout any cluster by making small local changes; i.e. changing the values of $o(n)$ variables in each step.  But solutions in two different clusters must differ globally; i.e. they differ on a linear number of variables.

While these hypotheses are, for the most part, not rigorously proven, they come from some substantial mathematical analysis.  They explain many phenomena, most notably why some random CSP's are algorithmically very challenging (eg.~\cite{aco,dg}).  Intuition gained from these hypotheses  has led to  some very impressive heuristics (eg. Survey Propogation\cite{sp,mz,bmz}, and the best of the random $r$-SAT algorithms whose performance has been rigorously proven~\cite{coalg}), and some remarkably tight rigorous bounds on various satisfiability thresholds~\cite{cop,cop2,cov,Coja}. Ding, Sly and Sun recently used an approach outlined by these hypotheses to prove the $k$-SAT conjecture~\cite{DSS}, with a determination of the $k$-SAT satisfiability threshold for all large $k$.  It is clear that, in order to approach many of the outstanding challenges around random CSP's, we need to  understand clustering.

Amongst the commonly studied random CSP's, $r$-XORSAT (a.k.a.\ linear equations over GF(2)) is the one for which the clustering picture is, rigorously, the most well-established. The exact satisfiability threshold was established for $r=3$ in \cite{dub}, and then for $r\geq 4$ in \cite{cuc,ps}. The clustering threshold $c_r^*$, and the structure of the solution clusters were analyzed in~\cite{cdmm,mez} and then established rigorously in \cite{ikkm, amxor}.
These papers  provide} a very thorough description of the clusters for any constant density $c\neq c_r^*$. However, the birth of the $r$-XORSAT clusters is not well understood. There has been no description of the solution space when the number of constraints is  $c_r^*n+o(n)$. This is the main target of this paper.

Consider a random process where random $r$-XORSAT constraints are added one at a time; this corresponds to a random hypergraph process where random $r$-uniform hyperedges are added one after the other. We show that the the geometric structure of the whole solution space,
specifically a key connectivity parameter, transits rather smoothly around clustering. Analysing the manner of a transition near the threshold of a phase transition is a common goal; see e.g.\  the extensive work on the birth of the giant component\cite{bb3,tlcomp,lpw,jlkp}, and the 2-SAT transition\cite{bbckw}.

The cluster structure of $r$-XORSAT is simpler than that of most other models, and this  has enabled researchers to prove challenging results for it long before proving them for other models. However, the structure of the clusters in the other models are hypothesized to be a generalization of the simpler structure in $r$-XORSAT. Understanding cluster properties in $r$-XORSAT often helps to predict properties in other CSP's. For instance, many CSP's have the generic property that, after the freezing threshold, most clusters consist of a solution $\s$ to a subset (of linear size) of the variables, called {\em frozen variables}, along with all extensions of $\s$ to the rest of the variables~\cite{art,mmfreeze,bmz,gs}.  This is true of $r$-XORSAT clusters, although in a simpler way: the set of frozen variables is invariant among clusters, whereas in other models, they can differ.   Insights from the work on $r$-XORSAT have been valuable when studying more complicated models; eg. ideas from~\cite{amxor} led to~\cite{mmfreeze,mres}.

\section{Main results}

Before stating the main results, we formalise the concepts of clustering and connectivity of clusters mentioned in Section~\ref{sec:intro}. We also give formal definitions of probability spaces under the discussion of this paper.

\subsection{Random linear equations and random hypergraphs}

Our model for a random system of equations is $X_r(n,m)$ defined as follows.  We have $n$ variables over GF(2) and {$m$ equations each formed in the following way: The LHS is the summation of a uniformly chosen $r$-tuple of variables and the RHS is chosen uniformly from $\{0,1\}$.}  We focus on the case $m=cn$ where $c=\Theta(1)$.  We restrict ourselves to the case $r\geq 3$, as the case $r=2$ behaves very differently, and is already well-understood (see eg.\cite{bbckw}).  For this range of $r,m$, a simple first moment bound shows that \aas\ no two equations will have the same $r$-tuple of variables, so  choosing the $r$-tuples with or without replacement has a negligible effect; to be specific, we choose them without replacement.

It is not hard to see that this is equivalent to choosing an instance of $r$-XORSAT on $n$ variables by {uniformly choosing $m$ $r$-tuples to form clauses}, and then signing the variables within each clause uniformly at random. An assignment to the variables is satisfying if every clause contains an odd number of true literals.

A common alternate model is to choose each of the $\binom{n}{r}$ $r$-tuples of variables independently with probability $p$, and then form an equation for each $r$-tuple (with a uniformly random RHS). By conditioning on the ``typical'' number of chosen $r$-tuples, our results for $X_r(n,m)$ immediately translates to this model.

Given a system of linear equations over GF(2), its {\em underlying hypergraph} is defined as follows: the vertices are the variables and for each equation, the set of variables appearing in
that equation form a hyperedge. 
So the underlying hypergraph of $X_r(n,m)$ is distributed as $\calh_r(n,m)$, the
random $r$-uniform hypergraph on $n$ vertices and $m$ hyperedges uniformly chosen without replacement from the $\binom{n}{r}$ hyperedges in the complete hypergraph.

\subsection{Clustering}

We formalise the concept of solution clustering in CSPs. Given a CSP instance $F$, let $\Phi(F)$ denote the set of solutions of $F$. Let $f=f(n)$ be an integer between $1$ and $n$. Construct a graph $G(\Phi(F),f)$ as follows. The set of vertices in $G(\Phi(F),f)$ is $\Phi(F)$; two vertices $x$ and $y$ are adjacent if their Hamming distance is at most $f$. Now,
we say two solutions $\sigma_1$ and $\sigma_2$ of $F$ are {\em $f$-connected} if $\sigma_1$ and $\sigma_2$ are in the same component of $G(\Phi(F),f)$; conversely, we say $\sigma_1$ and $\sigma_2$ are {\em $f$-separated} if they are not $f$-connected.
Given a set of solutions $S$, we say $S$ is {\em $f$-connected} if all solutions in $S$ are in the same component in $G(\Phi(F),f)$; i.e.\ all solutions in $S$ are pairwise $f$-connected. Given two disjoint sets of solutions $S_1$ and $S_2$, we say they are {\em $f$-separated} if for every $\sigma_1\in S_1$, $\sigma_2\in S_2$, $\sigma_1$ and $\sigma_2$ are $f$-separated.

The clustering phenomenon described in Section~\ref{sec:intro} basically says that given a particular CSP, there exist a constant $c^*>0$ and two functions $f(n)=o(n)$ and $g(n)=\Omega(n)$ such that, if the density of a random CSP instance $F$ is below $c^*$, then \aas\ all solutions of $F$ are $f(n)$-connected; if the density of $F$ is above $c^*$, then \aas\ the solutions can be partitioned into many clusters, each cluster corresponding to a component in $G(\Phi(F),f(n))$: every cluster is $f(n)$-connected whereas every pair of clusters are $g(n)$-separated. This is to say that one can walk from any solution to any other inside the same cluster by changing at most $f(n)$ variables at a time; but to walk from one solution to another one in a different cluster, one must change more than $g(n)$ variables in one step. It has been proved~\cite{amxor} that for a random $r$-XORSAT whose density is a constant not equal to $c^*_r$, $f(n)$ can be chosen as $C\log n$ for some sufficiently large constant $C>0$.

\subsection{Solution clusters}\lab{ssc}

In this section, we give a full characterisation of clusters of a random linear equation system (or equivalently $r$-XORSAT).

 Given a hypergraph $H$, its {\em $k$-core}, denoted by $\calc_k(H)$, is the {maximum} subgraph of $H$ in which every vertex has degree at least $k$. The $k$-core of a hypergraph can be easily obtained by removing repeatedly every vertex with degree less than $k$. 

  The {\em 2-core} of a system of linear equations is the subset of equations corresponding to the hyperedges that are in the 2-core of the underlying hypergraph. The $2$-core of $X_r(n,m)$ plays an essential role of determining the clustering threshold for and characterising the clusters of $X_r(n,m)$.
 It is easy to see that every solution to the 2-core of $X_r(n,m)$ (corresponding to the 2-core of $\calh_r(n,m)$) can be easily extended to a solution of the entire system, by setting the other variables in the reverse order that they are removed.

Roughly speaking, the clusters correspond to solutions of the  2-core.
But this is not quite true - we need to account for the effects of short {\em flippable cycles} which we define as follows:

\begin{definition}\label{flippable_def}
A \emph{flippable cycle} in a hypergraph {$H$} is a set of vertices $S=\{v_0,\ldots,v_t\}$
{where the set of hyperedges incident to $S$ can be ordered as $e_0, \ldots, e_t$ such that}
each vertex $v_i$ lies in $e_i$ and in $e_{i+1}$ and in no other {edges of $H$}
 (addition mod $t$).
\end{definition}
Thus, the vertices $v_0,\ldots,v_t$ must have degree exactly two in the hypergraph. The remaining vertices in hyperedges $e_0,\ldots,e_t$ can have arbitrary degree and are \emph{not} part of the  flippable cycle. 

\begin{definition}\label{coreflippable_def}  A \emph{core flippable cycle} in  a hypergraph $H$ is a flippable cycle in the subhypergraph  induced by the 2-core of $H$.
\end{definition}

Thus, in a core flippable cycle, the vertices $v_0,\ldots,v_t$ have degree exactly two \emph{in the 2-core}, but possibly higher degree in $H$. Note also that {$H$ may contain} flippable cycles outside the 2-core. If we take a solution $\s$ to the entire system, and change the assignment to each variable in a flippable cycle of the underlying hypergraph, we obtain another solution $\sigma'$. If the change is on a small core flippable cycle, then $\sigma$ and $\sigma'$ differ by a small number of variables and thus they should be in the same cluster, even though they do not agree on the set of variables in the 2-core. This suggests that we have to take core flippable cycles into consideration when we characterise the structure of the clusters. 

In Section~\ref{sec:cycle}, we will show that for the random hypergraphs studied in this paper, very few vertices lie in core flippable cycles.

\begin{definition}
Two solutions are \emph{cycle-equivalent} if on the 2-core they differ only on variables in core flippable cycles ({while} they may differ arbitrarily on variables not in the 2-core).
\end{definition}

\begin{definition}\label{cluster_def}
The {\em solution clusters} of {$X_r(n,m)$} are the cycle-equivalence classes, i.e., two solutions are in the same cluster iff they are cycle-equivalent.
\end{definition}

In other words: Let $\s$ be any solution to the subsystem induced by the 2-core.  It is easy to see that $\s$ can be extended to a solution to the entire system of equations.  All such extensions, along with all extensions of any 2-core solutions obtained by altering $\s$ only on core flippable cycles, form a cluster.
By symmetry, all clusters are isomorphic.
Note that, if the 2-core is empty, then our definitions imply that all solutions are in the same cluster. 
So the clustering threshold for $X_r(n,cn)$ corresponds to
the emergence threshold of a non-empty 2-core for $\calh_r(n,cn)$.

The threshold for the appearance of a non-empty $k$-core in $\calh_{\dd}(n,cn)$ ($(k,r)\neq (2,2)$) is determined\cite{psw,mmcore,jhk}, given as below.
  \begin{equation}\lab{krthreshold}
c_{\dd,k}=\inf_{\mu > 0}
 \frac{\mu}{r\left[e^{-\mu}\sum_{i = {k-1}}^{\infty} \mu^i/i!\right]^{\dd-1}}
 \enspace .
 \end{equation}
 Define
 \be\lab{cr}
 c_r^*=c_{r,2}.
 \ee
The following theorem confirms that the clustering threshold  for $X_r(n,cn)$ is $c_r^*$.

\begin{theorem}[\cite{amxor,ikkm}] For every fixed integer $\dd\geq 3$ and real number $\eps>0$, 
\begin{enumerate}
\item[(a)] if $c\le c_r^*-\eps$ then all solutions of $X_{\dd}(n,cn)$ are $O(\log n)$-connected;
\item[(b)] if $c\ge c_r^*+\eps$ then the solutions of $X_{\dd}(n,cn)$ are partitioned into well-connected well-separated clusters: every cluster is $O(\log n)$-connected and every pair of clusters are $\Omega(n)$-separated.
\end{enumerate}

\end{theorem}

\subsection{Our contribution}

Recall the definition of solution clusters of $X_r(n,cn)$ in Definition~\ref{cluster_def} and the clustering threshold in~\eqn{cr}. We prove that when $c\to c_{r}^*$, the connectivity of clusters of $X_r(n,cn)$ undergoes a smooth transition.  In the following theorem, we describe the solution geometry of $X_r(n,cn)$ for $c=c_r^*+n^{-\d}$ when $\delta>0$ is sufficiently small.

\begin{theorem}\lab{tc1}  Fix constant $\dd\geq 3$. There exist constants $\k=\k(r), Z=Z(r)>0$  such that for any sufficiently small constant $\d>0$, in $X_{\dd}(n,cn)$ where $c=c_r^*+n^{-\d}$, \aas:
\begin{enumerate}
\item[(a)] every cluster is $n^{\k\d}$-connected;
\item[(b)] every pair of clusters are $Zn^{1-r\d}$-separated;
\item[(c)]  there exists a pair of solutions $\s,\t$ in the same cluster such that $\s$ and $\t$ are $n^{\d/20}$-separated.
\end{enumerate}
\end{theorem}
\no {\bf Remarks}.

\no(i) The arguments below will imply that in the setting of Theorem~\ref{tc1} there are many clusters.  Lemma~\ref{lcoresize} yields that the number of variables in the 2-core exceeds the number of equations on those variables by $\Theta(n)$.  Lemma~\ref{lflip} says that $o(n)$ of these variables are on flippable cycles. It easily follows that there are at least $e^{\Theta(n)}$ clusters.

\no(ii) Theorem~\ref{tc1}(b) does not exclude the possibility that different clusters are, in fact, linearly separated; i.e. that there is some constant $\a>0$ such that \aas\ every pair of clusters are $\a n$-separated.

\no(iii) In fact, we prove that
Theorem~\ref{tc1}(a,b) holds for all $\d<\hf$, but note that these results are trivial for  $\d\geq\inv{\k}$ (part (a)) and $\d\geq\inv{r}$ (part (b)).

\no(iv) Theorem~\ref{tc1}(c) shows that
Theorem~\ref{tc1}(a) is best possible, up to the value of $\k$.

\smallskip

For $c=c_r^*-n^{-\d}$, we prove that all solutions are contained in a single cluster that is $n^{O(\d)}$-connected.

\begin{theorem}\lab{tc3}  For $\dd\geq 3$,  there exists $\k=\k(r)$  such that: for any $0<\d<1/2$ and $c=c_r^*-n^{-\d}$,  \aas\ all solutions of  $X_{\dd}(n,cn)$ are  $n^{\k\d}$-connected.
\end{theorem}

We conjecture that the above theorem is tight up to the value of $\k$, just like Theorem~\ref{tc1}(a) is. However, the proof of Theorem~\ref{tc1}(c) does not generalise to $c=c_r^*-n^{-\d}$.
\begin{conjecture}\lab{conj} For $\dd\geq 3$,  there exists $\k'=\k'(r)$  such that: for any sufficiently small $\delta>0$ and $c=c_r^*-n^{-\d}$, a.a.s.\ there exist two solutions $\s,\t$ of $X_{\dd}(n,cn)$ such that $\s$ and $\t$ are $n^{\k'\d}$-separated.
\end{conjecture}

\begin{figure}[h]\label{fig1}
\begin{center}
 \includegraphics[width=0.9\textwidth]{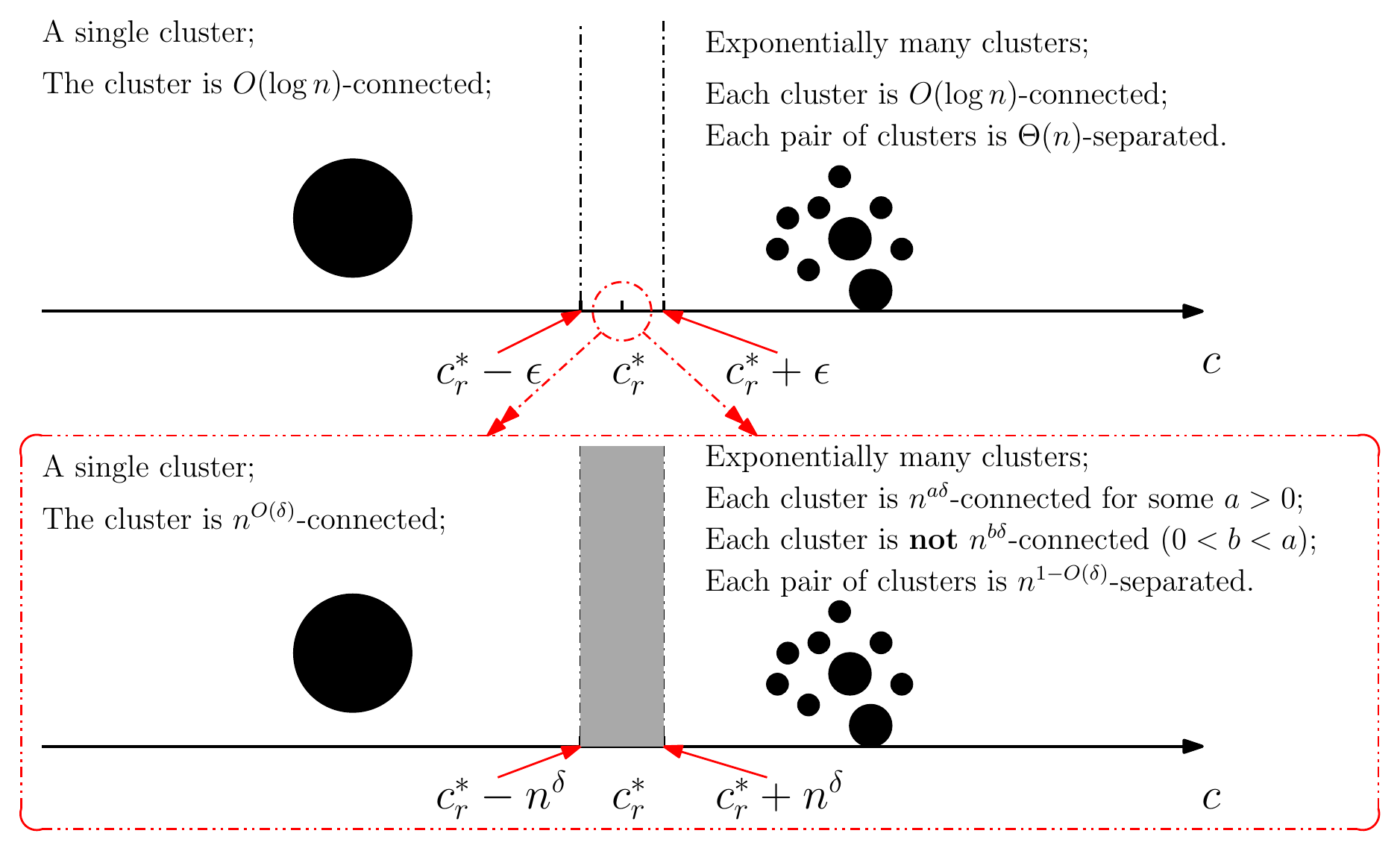}
 \caption{The top figure shows what was previously known~\cite{amxor,ikkm}.  The bottom figure shows the results of this paper.}
\end{center}
\end{figure}

Our results are summarized in Figure~\ref{fig1}.

Theorem~\ref{tc1} requires most of the new ideas in this paper. The analogous result in~\cite{amxor} is Observation 10, a simple observation regarding high degree vertices.  To prove this result, we need to dig into the 2-core stripping process and show that a variable $u$ removed late in the process has the following property: any two solutions that disagree on $u$ must disagree on at least $n^{\Theta(\delta)}$ other vertices; in fact, we identify those other vertices. These arguments appear in Section~\ref{sec8}.
The arguments for our other results are similar to those in~\cite{amxor}. But where the arguments in~\cite{amxor} made use of facts about the 2-core of a hypergraph with density a constant $c>c_r^*$, this paper requires analogous results for the much more difficult range $c=c_r^*+o(1)$.  Those results were derived in~\cite{gm,gm2}; the results of this paper were the motivation for those two papers.

\section{Size of the 2-core}
\lab{sec:core}
Our analysis relies on the size of the 2-core of $\calh_{r}(n,cn)$. This (indeed the $k$-core for every $k\ge 2$) has been well studied in, eg.\ ~\cite{psw,mmcore,jhk, fr}.  Here we present the well-established expressions for the numbers of vertices and edges.
For any  real $\la>0$ and integer $t\geq 0$ we define $f_t(\la)$ to be the probability that a Poisson variable with mean $\la$ is at least $t$; i.e.
\begin{eqnarray*}
f_{t}(\la)&=&e^{-\la}\sum_{i\geq t}\frac{\la^i}{i!}.\\
\end{eqnarray*}

Fix $\dd,k\geq2, (\dd,k)\neq (2,2)$.  For any $c>0$ we define:
\[h(\mu)=h_{\dd,k}(\mu)=\frac{\mu}{f_{k-1}(\mu)^{\dd-1}};\]
and for any $c\ge c_{r,k}$, we define:
\[\mu(c) \mbox{ is the larger solution to }
c=h(\mu)/r.\]
Note that~(\ref{krthreshold}) defines $c_{r,k}$ to be the minimum value of $c$ such that $\mu(c)$ is positive.

One can show that the degree sequence of the $k$-core is distributed (approximately) like a Poisson with mean $\mu(c)$ truncated at being at least $k$.  Indeed, the probability that a vertex $v$ of our random graph is in the $k$-core is (roughly) the probability that a Poisson with mean $\mu(c)$ is at least $k$.  This, and simple calculations, show that the expected number of vertices and edges in the $k$-core are $\a(c) n +o(n)$ and $\b(c)n+o(n)$ where:
\begin{eqnarray*}
\a(c)&=&f_k(\mu(c)),\quad \b(c)=\frac{1}{r}\mu(c) f_{k-1}(\mu(c)).
\end{eqnarray*}
(See any of~\cite{psw,mmcore,jhk,fr} for details.) Since we are focussing on the $k$-core when $c=c_{r,k}+o(1)$, we define
\be
\mu_{\dd,k}=\mu(c_{\dd,k}),\quad
\a_{\dd,k}=f_k(\mu_{\dd,k}),\quad
\b_{\dd,k}=\frac{1}{\dd}\mu_{\dd,k} f_{k-1}(\mu_{\dd,k}). \lab{alpha-beta}
\ee
 For ease of notation, we drop most of the $\dd,k$ subscripts.

We will use the following result by Kim~\cite{jhk} (for $k=2$). 

\begin{theorem}\lab{tkim} Fix $\dd,k\geq2, (\dd,k)\neq (2,2)$ and  an arbitrary constant $\e>0$.
\begin{enumerate}
\item[(a)] For $c\le c_{\dd,k}-n^{-1/2+\e}$, \aas\ the $k$-core  of $\calh_{\dd}(n,cn)$ is empty.
\item[(b)] For $c\ge c_{\dd,k}+n^{-1/2+\e}$, \aas\ $\calh_{\dd}(n,cn)$ has a $k$-core with $\a(c)n+O(n^{3/4})$  vertices and $\b(c)n+O(n^{3/4})$ hyperedges.
\end{enumerate}
\end{theorem}

With some elementary calculations on $\a(c)$ and $\b(c)$ when $c=c_{r,k}+o(1)$, the following lemma follows immediately from Theorem~\ref{tkim}. The detailed calculations can be found in~\cite[Lemma 8]{gm}.
\begin{lemma}\lab{lcoresize}
For  $\dd,k\geq2, (\dd,k)\neq (2,2)$, there exist {three positive constants} $K_1=K_1(r,k),K_2=K_2(r,k),K_3=K_3(r,k)$ such that: if $c= c_{\dd,k}+ n^{-\d}$ for some constant $0<\d<1/2$, then \aas\ the $k$-core of $\calh_{\dd}(n,cn)$ has
$\a n +K_1 n^{1-\d/2} +O(n^{1-\d}+n^{3/4})$ vertices, 
$\b n +K_2 n^{1-\d/2} +O(n^{1-\d}+n^{{3/4}})$ hyperedges, and  average degree $r\b/\a+K_3n^{-\d/2}+O(n^{-\d}+n^{-1/4})$. \end{lemma}

\section{AP-model and degree distribution of the $k$-core}
\lab{sec:AP}


As usual, we will analyze the $k$-core using a model which allows us to capture the degree sequence.  We will use the AP-model (the allocation and partitioning model), first introduced in~\cite{CW}. 

We start with $n$ distinct bins and $rm$ vertex-copies. The probability space generated by the AP-model, denoted by $AP_r(n,m)$ can be described as follows: Allocate each vertex-copy uniformly at random to one of the $n$ bins; take a uniform partition of the $rm$ vertex-copies into $m$ parts; each of size $r$. The resulting is a configuration, which is a random element in $AP_r(n,m)$. 
  Representing each bin as a vertex and each $r$-tuple in the partition as a hyperedge, a configuration in $AP_r(n,m)$ corresponds to a multihypergraph on $n$ vertices and $m$ hyperedges.  
  
Of course, a configuration in $AP_r(n,m)$ does not necessarily correspond to a simple hypergraph, but an easy counting argument shows that every simple hypergraph in $\H_r(n,m)$ corresponds to the same number of configurations in $AP_r(n,m)$ and thus, $\H_r(n,m)$ is the random hypergraph generated by $AP_r(n,m)$, conditional on being simple. When $m=O(n)$, the probability that $AP_r(n,m)$ generates a simple hypergraph is $\Theta(1)$~\cite{Chvatal}. Thus we immediately have the following corollary.
  
  \begin{corollary}\lab{cor:AP}
 For any $m=O(n)$: If $A_n$ is an event that a.a.s.\ holds in $AP_r(n,m)$, then $A_n$ holds a.a.s.\ in $\H_r(n,m)$.
  \end{corollary}

Note that all bounds from Section~\ref{sec:core} on the size of the 2-core hold for the AP-model (indeed these bounds can be obtained by analysing the AP-model).  We will also use the following result on the degree sequence of the $k$-core. A proof can be found in~\cite[Corollary 2]{CW}. 

\begin{proposition}\lab{p:Poisson} Let $\zeta$ denote the average degree of the $k$-core of $AP_r(n,cn)$.
For any constant $j\ge k$, let $\rho_j$ denote the proportion of vertices in the $k$-core whose degree equals $j$. Then, for any $\eps>0$, a.a.s.\
\begin{equation}
\rho_j=e^{-\la}\frac{\la^j}{f_k(\la)j!}+O(n^{-1/2+\eps}), \lab{PoissonApprox}
\end{equation}
where $\la$ satisfies $\la f_{k-1}(\la)/f_k(\la)=\zeta$.
\end{proposition}

When $c$ is very close to $c_{r,k}$, many parameters in the $k$-core of $AP_r(n,cn)$ (if it is not empty) are very close to certain critical values. For instance, it is easy to see that the average degree of the $k$-core is very close to 
\[
\zeta:=\frac{r\beta}{\alpha}=\mu_{r,k}\frac{f_{k-1}(\mu_{r,k})}{f_k(\mu_{r,k})}.
\] The following equation has appeared in several prior papers relating to the analysis the $k$-core. For instance, it is displayed in~\cite[Eq.\ (11)]{gm}. We will use this equation in Section~\ref{sec:cycle}.
\be
e^{-\mu_{\dd,k}}\frac{\mu_{\dd,k}^{k-1}}{(k-2)!f_{k-1}(\mu_{\dd,k})}=\frac{1}{(\dd-1)}.\lab{degreeK}
\ee
\remove{
The AP-model differs from the configuration model, first introduced by Bollob\'{a}s~\cite{bb}, in the sense that the degree sequence is determined by random allocation of vertex-copies instead of being specified initially. If we specify a degree sequence ${\bf d}$ and conditioning on that the random allocation produces ${\bf d}$, then the AP-model generates an equivalent probability space of that by the configuration model. When we apply the configuration model, rather than allocating vertex-copies to bins, we start with vertex $i$ as a bin containing $d_i$ vertex-copies; then take a uniform partition of all vertex-copies such that each part has size exactly $\dd$. The configuration model is more convenient to use if the degree sequence is exposed.
\begin{definition}\lab{def:niceDeg}
A {degree sequence {\bf d}} is \emph{nice} if $\sum_{1\le i\le n} d_i=\Theta(n)$, $\sum_{1\le i\le n} d_i^2=O(n)$
and $d_i=o(n^{1/24})$ for all $i$.
\end{definition}

The following  proposition is of a standard way of translating results in $\M({\bf d})$ to the uniform model.

\begin{proposition}[\cite{cc}] Assume that  ${\bf d}$ is {a nice degree sequence.} if property $Q$ holds a.a.s.\ in $\M({\bf d})$, then $Q$ holds a.a.s.\ for a uniformly random simple hypergraph with degree sequence ${\bf d}$.
\end{proposition}

 We can easily verify that a.a.s.\ all the degree sequences considered in this paper are nice.
}

\section{Bounding core flippable cycles}\lab{sec:cycle}

As core flippable cycles influence the clusters of $X_r(n,cn)$, we prove in this section that not many vertices lie on core flippable cycles.

\begin{lemma}\lab{lflip} Fix $r\geq 3$. For any $0<\d<\hf$ and $c=c_{\dd,2}+n^{-\d}$, \aas\
the total sizes of all core flippable cycles in $\H_r(n,cn)$ is at most $O(n^{\d/2}\log n)$.
\end{lemma}

\proofstart
We follow a similar analysis to that in~\cite[Lemma~35]{amxor}.  We work in the AP-model;  Corollary~\ref{cor:AP} then implies that the result holds for $\H_r(n,cn)$. 
 Recall the definitions of $\mu(c)$, $\alpha(c)$ and $\beta(c)$ from the beginning of Section~\ref{sec:core}. By Theorem~\ref{tkim}, \aas\ the $2$-core contains $Q$ vertices with total degree $\La$ where
 \be
  Q=\alpha(c) n+ O(n^{3/4}),\quad  \La=\dd\beta(c)n+O(n^{3/4}).\lab{Q}
  \ee 
  Let $Q_2$ denote the number of vertices in the 2-core with degree $2$. 
We will prove below that
\be \lab{eq2}
\frac{2(r-1)Q_2}{\La}\leq 1-Kn^{-\d/2},
\ee
for some constant $K>0$.

For any $a\geq1$, we let $X_a$ denote the number of core flippable cycles of size $a$. The calculations from the proof of Lemma 35 of~\cite{amxor}, which are in fact a simple exercise, say:

\[
\ex X_a\le {Q_2\choose a}\frac{(a-1)!}{2}2^a\prod_{\ell=1}^a\frac{\dd-1}{\La-2\ell+1}
<\frac{1}{2a}\prod_{\ell=1}^a\frac{(\dd-1)(2Q_2-2\ell+2)}{\La-2\ell+1}.
\]

Since $r-1\ge 2$, we have $2Q_2/\La\le \hf$ and so $\frac{2Q_2-2\ell+2}{\La-2\ell+1}\leq \frac{2Q_2}{\La}$ for each $\ell$. So~(\ref{eq2}) yields:

\begin{equation}\label{efc1}
\ex X_a\le \frac{1}{2a}\prod_{\ell=1}^a\frac{2(\dd-1)Q_2}{\La}\leq \frac{1}{2a}(1-Kn^{-\d/2})^{a}.
\end{equation}
Now let $X$ denote the total number of vertices appearing on core flippable cycles.
\[\ex X \leq \ex \sum_{a\geq 1}aX_a < \sum_{a\geq 1}(1-Kn^{-\d/2})^{a}<\inv{K}n^{\d/2}.\]
So Markov's Inequality yields $\pr(X>n^{\d/2}\log n)=o(1)$ which proves the lemma.

It only remains to prove~(\ref{eq2}). 
By Proposition~\ref{p:Poisson},
 for any $\e>0$, a.a.s.
\[
\frac{Q_2}{Q}=e^{-\la} \frac{\la^2}{2f_2(\la) }+O(n^{-1/2+\eps}),
\]
where $\la$ satisfies
$\la f_1(\la)/f_2(\la)=\La/Q$.  
Conditional on any values of $Q$ and $\La$ {satisfying~\eqn{Q},} we have 
 \[
 \la \frac{f_1(\la)}{f_2(\la)}=\frac{\dd\beta(c)}{\alpha(c)}+O(n^{-1/4}).
 \]
The function $g(x)=x f_1(x)/f_2(x)$ is a strictly increasing function on $x>0$ (see~\cite[Lemma 26]{gm} for a proof). Then, by the definition of $\a(c)$, $\b(c)$ and $\mu(c)$ above~\eqn{alpha-beta}, we have
\[
\la=\mu(c)+O(n^{-1/4}).
\]
Immediately, we have
\be\lab{er2}
Q_2=\left(\frac{e^{-\mu(c)} \mu(c)^2}{2f_2(\mu(c))}+O(n^{-1/4})\right) Q=\left(\frac{e^{-\mu(c)} \mu(c)^2}{2f_2(\mu(c))}+O(n^{-1/4})\right)\a(c) n,
\ee
since the function $e^{-\la} \la^2/2f_2(\la)$ has bounded derivative at $\la=\mu(c)$ and the error $O(n^{-1/2+\eps})$ is absorbed by $O(n^{-1/4})$. 
By~\eqn{degreeK},
\[
e^{-\mu_{\dd,2}}\frac{\mu_{\dd,2}}{f_1(\mu_{\dd,2})}=\frac{1}{(\dd-1)}.
\]
It is easy to check that the derivative of
$e^{-\mu}\frac{\mu}{f_1(\mu)}$ with respect to $\mu$ is strictly negative in a small neighbourhood of $\mu_{\dd,2}$. By Lemma~\ref{lcoresize}, $\mu(c)= \mu_{r,2}+ K_2n^{-\d/2}+o(n^{-\d/2})$ for some constant $K_2>0$. Hence,
\be
e^{-\mu(c)}\frac{\mu(c)}{f_1(\mu(c))}=e^{-\mu_{\dd,2}}\frac{\mu_{\dd,2}}{f_1(\mu_{\dd,2})}-K_3n^{-\d/2}+o(n^{-\d/2})=\frac{1}{(\dd-1)}-K_3n^{-\d/2}+o(n^{-\d/2}),\lab{err3}
\ee
for some constant $K_3>0$.  
      Now, by~\eqn{Q} and~\eqn{er2}, and recalling the definition of $\a(c)$ and $\b(c)$ above~\eqn{alpha-beta},
\bean
\frac{2(r-1)Q_2}{\La}&=&\left(e^{-\mu(c)}\frac{\mu(c)^2}{2f_2(\mu(c))}+O(n^{-1/4})\right)\frac{2(r-1)\a(c)}{r\b(c)}\\
&=&\left(e^{-\mu(c)}\frac{\mu(c)^2}{2f_2(\mu(c))}+O(n^{-1/4})\right) 2(r-1) \frac{f_2(\mu(c))}{\mu(c) f_1(\mu(c))}\\
&=&(r-1)\cdot e^{-\mu(c)}\frac{\mu(c)}{f_1(\mu(c))}+O(n^{-1/4})\\
&=&1-K_3(r-1)n^{-\d/2}+O(n^{-1/4})+o(n^{-\d/2}) \quad \mbox{by~\eqn{err3}}.
\eean
Since $\d<1/2$, we have $n^{-1/4}=o(n^{-\d/2})$. Then~(\ref{eq2}) follows.
 \proofend

Next we show that \whp\ the core flippable cycles are disjoint:

\begin{lemma}\label{ldfc} Fix $r\geq 3$. For any $0<\d<\hf$ and $c=c_{\dd,2}+n^{-\d}$, \aas\ no vertex lies in two different core flippable cycles.
\end{lemma}

\proofstart If a vertex $v$ lies in two core flippable cycles, then we must have the following structure (see {a detailed description in} the proof of Lemma 35 in the arXived version of \cite{amxor}, which uses different notation):   One core flippable cycle with vertices $v=v_1,...,v_a$ where each pair $v_i,v_{i+1}, i=1,...,a$ shares a hyperedge; and a sequence of vertices $u_1,...,u_q, q\leq a$ {(they form part of the other core flippable cycle)} where (i) each pair $u_i,u_{i+1}, i=1,...,q-1$ shares a hyperedge which contains none of $v_1,...,v_a$, (ii) $u_1$ is in the hyperedge containing $v_j,v_{j+1}$ and (iii) $u_q$ is in the hyperedge containing $v_{j'},v_{j'+1}$ for some $j'\neq j$.  All these vertices have degree two.

Calculations very similar to those in the previous proof bound the expected number of such structures for a given value of $a=o(n)$ as follows.  The $a^2$ term comes from $a$ choices for each of $j,j',q$ multiplied by the $\inv{2a}$ term from~(\ref{efc1}); the $\inv{\Lambda}$ term comes from the fact that the double cycle produces $a+q+1$ different $\Theta(\inv{\Lambda})$ terms but only $a+q$  different $\Theta(n){=\Theta(\Lambda)}$ terms.
\[O\left(\frac{a^2}{\Lambda}\right)(1-Kn^{-\d/2})^{a+q}.\]
Lemma~\ref{lflip} allows us to restrict to $a=O(n^{\d/2}\log n)$.  Summing over all such $a$   yields that the expected number of these structures is at most
\[O(1)\frac{(n^{\d/2}\log n)^3}{\Lambda}(1-Kn^{-\d/2})^{a}=O(n^{3\d/2-1}\log^3 n)=o(1),\] 
for $\d<\hf$.
\proofend

\section{Proof of Theorems~\ref{tc1}(a) and~\ref{tc3}}\lab{sec:free}

Recall that the $k$-core of a hypergraph $H$ can be obtained by repeatedly removing vertices with degree less than $k$.

\begin{definition}\lab{def:depth} A {\em $k$-stripping sequence} is a sequence of vertices that can be deleted from a hypergraph, one-at-a-time, {along with their incident hyperedges} such that at the time of deletion each vertex has degree less than $k$.
\end{definition}

Let $H$ be an $r$-uniform hypergraph and let $\Psi=v_1v_2\cdots$ be a $k$-stripping sequence which contains all non-$k$-core vertices of $H$. We create a directed graph (not a directed hypergraph) $\cald(\Psi)$ associated with $\Psi$ as follows. The vertices in $\cald(\Psi)$ are a subset of the vertices of $H$ -- specifically, the vertices not in the $k$-core and the vertices of the $k$-core that have a neighbour outside of the $k$-core. At the moment when $v_i$ is to be deleted from $H$, consider each hyperedge $x$ that is incident with $v_i$ (there are at most $k-1$ of them).
Add a directed edge to $v_i$ in $\cald(\Psi)$ from each of the $r-1$ other vertices in $x$.

For any vertex $v\in \cald(\Psi)$, we define $R_{\Psi}^+(v)$ to be the set of vertices reachable from $v$ in $\cald(\Psi)$. 
We have the following bound on $|R_{\Psi}^+(v)|$. Part (a) is from~\cite[Theorem 43]{gm} (for $c=c_{r,k}+n^{-\d}$) and~\cite[Theorem 6]{gm2} (for $c=c_{r,k}-n^{-\d}$), and part (b) is from~\cite[Theorem 5]{gm} (for $c=c_{r,k}+n^{-\d}$) and~\cite[Theorem 4]{gm2} (for $c=c_{r,k}-n^{-\d}$).
\begin{theorem}\lab{mt2}
Let $\dd,k\geq2, (\dd,k)\neq (2,2)$ be fixed. There is a constant $\kappa>0$ such that: for any constant $0<\d<\hf$,
if $c=c_{\dd,k}\pm n^{-\d}$, then 
\begin{enumerate}
\item[(a)]\aas\
there is a $k$-stripping sequence $\Psi$ containing all non-$k$-core vertices of $\calh_r(n,cn)$ such that $|R_{\Psi}^+(v)|\le n^{\kappa\d}$ for all $v\in\Psi$.
\item[(b)] a.a.s.\ for every $k$-stripping sequence $\Psi$, there exists a non-$k$-core vertex $v$ for which $|R_{\Psi}^+(v)|=\Omega(n^{\d/2})$.
\end{enumerate}
\end{theorem}

For the purposes of studying $r$-XORSAT clusters, we will only apply Theorem~\ref{mt2} for $k=2$. We use the stripping sequence $\Psi$ guaranteed by Theorem~\ref{mt2}(a).

 The argument
that our upper bound on $|R_{\Psi}^+(v)|$ implies that the clusters are well-connected is the same as that used in~\cite{amxor}.  We include it here for exposition, and because it is needed to understand the proof of Theorem~\ref{tc1}(c).

Each cluster is specified by an assignment to all of the 2-core variables not in any core flippable cycles; the cluster consists of all extensions from that assignment to the remaining variables.  More specifically, choose any such assignment $\s$ which does not violate any equations, substitute the value $v=\s(v)$ for each $v$ in the range of $\s$. This removes some of the linear equations (those whose variables are all set) and removes variables from some others.  The cluster now consists of all solutions to this reduced system.

If we apply Gaussian elimination to the equations of the reduced system, beginning with those in the core flippable cycles and then proceeding in the reverse order in which the equations were removed by $\Psi$, then we will obtain a system of equations in which each variable is expressed as the sum of a subset of what we call ${\cal F}$,  the {\em free variables}. It is a simple exercise (recall Lemma~\ref{ldfc} and see~\cite{amxor} for more details) to show that the free variables are as follows:

\begin{observation} ${\cal F}$ consists of the variables corresponding to: (i) the non-2-core vertices with indegree zero in $\cald(\Psi)$; (ii)  one vertex from each core flippable cycle.
\end{observation}


For each variable $v\notin\calf$ of the reduced system, there will be a set of free variables $\chi(v)\subseteq \calf$, such that the Gaussian elimination leaves exactly one equation containing $v$, and it is of the form
\be\lab{exv}
v=z_v+\sum_{u\in\chi (v)} u,
\ee
where $z_v$ is determined by $\s$ and thus is fixed for each cluster.
Furthermore, each variable of $\chi(v)$ can reach $v$ in $\cald(\Psi)$. (Note that $\chi(v)$ does not neccessarily contain every free variable that can reach $v$ in $\cald(\Psi)$.) Every equation is of this form, where $v\notin \calf$. So for each $u\in\calf$ we set $\chi(u)=\{u\}$ and $z_u=0$. Each of the $2^{|\calf|}$ possible assignments to the free variables is permissable.

It is important to stress that the set of free variables, and futhermore the sets $\chi(v)$, are determined only by our application of Gaussian elimination to the hypergraph, and not by any particular solution. So these are the same for every cluster.

Each cluster contains exactly $2^{|\calf|}$ solutions, one for each assignment to the free variables.  We can move from any solution to any other solution by changing the free variables, one-at-a-time, and updating the non-free variables using~(\ref{exv}). In the step where we change the free variable $u$, we only change the variables in
\[\chi^{-1}(u)=\{v: u\in\chi(v)\}.\]
If $u$ is not in the 2-core, then $\chi^{-1}(u)\subseteq R_{\Psi}^+(u)$ and so Theorem~\ref{mt2}(a) implies that we change at most $n^{\kappa\d}$ variables.   If $u$ is a free variable on a core flippable cycle $C$, then $\chi^{-1}(u)$ is contained in $\cup_w R_{\Psi}^+(w)$ over all $w\in \cald(\Psi)\cap C$. (Recall that the core flippable cycles are disjoint by Lemma~\ref{ldfc}.)  It is easy to show that a.a.s.\ the maximum degree of $\calh_r(n,cn)$ is less than $\log n$, and so Theorem~\ref{mt2}(a) implies $|R_{\Psi}^+(w)|\leq n^{\kappa\d} r\log n $ for each $w$ in a core flippable cycle.   By Lemma~\ref{lflip}, \aas\ the number of vertices in core flippable cycles is at most $n^{\d/2}\log n$ and so $|\chi^{-1}(u)|\leq n^{\kappa\d} r\log n  \times n^{\d/2}\log n<n^{(\kappa+1)\d}$. This proves
Theorems~\ref{tc1}(a) and~\ref{tc3}.
\proofend

\section{Proof of Theorem~\ref{tc1}(b)}

In this section, we describe the proof that solutions in different clusters are
$\Theta(n^{1-r\d})$-separated. As the proof is very similar to that of~\cite[Theorem 2]{amxor}, we only sketch the differences. 

\begin{proof}[Proof of Theorem~\ref{tc1}(b) (sketch)]
This follows the same argument as the proof of Theorem 2 of~\cite{amxor}.  The only change is to  Lemma~51 of~\cite{amxor}, where instead of proving that \aas\ there is no non-empty linked set  (see~\cite{amxor} for definitions) of size less than $\a n$, we prove that \aas\ there is none of size less than $n^{1-r\d}$.
(Caution:  in~\cite{amxor} the usage of $k,r$ is inverted from that of this paper.)

As in~\cite{amxor}, we use $X_a$ to denote the number of linked sets $S$ with $|\Gamma(S)|=a$ (see the definition of $\Gamma(S)$ in~\cite{amxor}).
Property (iii) at the beginning of the proof of Lemma~51, is equivalent to saying $\frac{2(r-1)Q_2}{\La}\leq 1-\z$, for some $\z>0$, in the notation of this paper.  Instead, we have $\frac{2(r-1)Q_2}{\La}\leq 1-Kn^{-\d/2}$ by (\ref{eq2}).  This results in replacing $Z_1=\Theta(1)$ from the proof of Lemma~51 of~\cite{amxor} with $Z_1=\Theta(n^{\d/2})$ (in the notation of~\cite{amxor}); this yields
\[\ex(X_a)<\left(\frac{\Theta(an^{r\d})}{n}\right)^{a/2r},\]
and it then follows easily that $\ex\left(\sum_{a=1}^{Zn^{1-r\d}}X_a\right)=o(1)$
 for sufficiently small $Z=Z(r)>0$. This proves the theorem.
\end{proof}

\section{Proof of Theorem~\ref{tc1}(c)}\lab{sec8}
We will make use of the characterization of the solution space in terms of free variables introduced in Section~\ref{sec:free}.  We will define a 2-stripping sequence $\Psi$, which in turn will define a set $\calf$ of free variables.  Recalling that each cluster contains one solution for each of the $2^{|\calf|}$ settings of the variables of $\calf$, it will suffice to show that there is at least one free variable $u$ such that any step which changes the value of $u$ will also change at least $n^{\d/20}$ other variables.

Recall that if we change a free variable $u$, and no other free variables, then we also change all the variables in $\chi^{-1}(u)\subseteq R_{\Psi}^+(u)$. Theorem~\ref{mt2} says that there is at least one variable $u$ with $|R_{\Psi}^+(u)|\geq n^{\Theta(\d)}$.   However, this does not immediately imply that each solution cluster is
not $n^{\Theta(\d)}$-connected.  For one thing, we require that there is a {\em free} variable $u$ such that  $R_{\Psi}^+(u)$ is that large.  In fact, we actually require $\chi^{-1}(u)\subseteq R_{\Psi}^+(u)$ to be that large. But even that would not suffice, since it would only imply that a step where $u$ is the only free variable changed would require changing $n^{\Theta(\d)}$ other variables.  It still leaves open the possibility that one could change $u$ using a step that also changes other free variables $w_1,...,w_t$ where $\cup\chi^{-1}(w_i)$ intersects $\chi^{-1}(u)$ and so not every variable in $\chi^{-1}(u)$ is changed.

To prove Theorem~\ref{tc1}(c) we prove that for our choice of $\Psi$, we have:

\begin{property} \lab{pfvu}
There is a free variable $u$, along with $n^{\d/20}$ other variables $v_1,...v_{n^{\d/20}}$ such that for each $i$, $\chi(v_i)=\{u\}$.
\end{property}

In order to move through every solution in a cluster, eventually there must be a step where $u$ is changed.  At that step, each of $v_1,...v_{n^{\d/20}}$ are changed as well, regardless of which other free variables are changed.  So the cluster is not $n^{\d/20}$-connected.

In order to find such $u$ and $v_1,\ldots, v_{n^{\d/20}}$, we need to specify a stripping sequence $\Psi$.  We start with defining a parallel stripping process which produces the $k$-core.

\begin{definition} \lab{parallel} The {\em parallel $k$-stripping process}, applied to a hypergraph $H$, consists of iteratively removing {\em all} vertices of degree less than $k$ at once along with any hyperedges containing any of those vertices, until no vertices of degree less than $k$ remain.
Let $S_i$ denote the set of vertices that are removed during iteration $i$. We use $\widehat{H}_i$ to denote the hypergraph remaining after $i-1$ iterations, i.e. after removing $S_1,...,S_{i-1}$.
\end{definition}

Note that the parallel $k$-stripping process terminates with the $k$-core of $H$. In order to define $\Psi$, we consider a slowed-down version of the parallel stripping process:

\bigskip

\noindent{\bf SLOW-STRIP:}  {\em We maintain  a queue $\Q$. Initially, $\Q$ is the set of all light vertices (vertices with degree less than $k$) of $H$. In each step of SLOW-STRIP, a hyperedge $x$ incident with the light vertex in the front of $\Q$ is removed. If any vertex becomes light after the removal of $x$, add it to the end of $\Q$. Repeatedly remove the vertex in the front of $\Q$ if its degree is zero. }

\bigskip

Note that the removal of $x$ might cause the degree of a vertex not at the front of $\Q$ to drop to zero.  That vertex remains in $\Q$ until it reaches the front, at which point it will be removed.  So it it possible that multiple vertices are removed from the front of $\Q$ during one step of SLOW-STRIP.

Let $\Psi$ be the stripping sequence produced by SLOW-STRIP and let $\cald=\cald(\Psi)$ be the digraph associated with $\Psi$ (recall the definition of $\cald$ below Definition~\ref{def:depth}).
Let $I^*$ be the largest value of $i$ such that $S_i$ contains a vertex with indegree zero in $\cald$ (recall that a vertex with indegree zero in $\cald$ is a free variable), and let $u^*\in S_{I^*}$ be some such vertex.

Our first step is to prove that there are no other free vertices within $n^{\d/20}$ levels of $u^*$:

\begin{lemma}\lab{lgap} Assume that $\delta>0$ is sufficiently small. A.a.s\ $u^*$ is the only free vertex in
$\cup_{i\geq I^*-n^{\d/20}} S_i$.
\end{lemma}

Given a non-core vertex $w$, we define $T(w)$ to be the set of vertices $v$ that can reach $w$ in $\cald$; i.e. the set of vertices $v$ such that $w\in R_{\Psi}^+(v)$. For $u\in T(w)$, define $T(w,u)$ to be the subgraph of $T(w)$ containing all vertices reachable from $u$; i.e. vertices on walks from $u$ to $w$.  We will prove:

\begin{lemma}\lab{lTw} Assume that $\delta>0$ is sufficiently small. A.a.s. there is some $w\in S_{I^*-n^{\d/20}}$ such that
\begin{enumerate}
\item[(a)] $u^*\in T(w)$;
\item[(b)] $|T(w)|\le n^{3\d/2}$; 
\item[(c)] the subgraph of $\cald$ induced by the vertices of $T(w,u^*)$ is a directed path, containing exactly one vertex in each $S_i$, $I^*-n^{\d/20}\le i\le I^*$.
\end{enumerate}
\end{lemma}

We defer the proofs of Lemmas~\ref{lgap} and~\ref{lTw} to Section~\ref{twoLemmas}.
\begin{proof}[Proof of Theorem~\ref{tc1}(c)] 

Choose a vertex $w\in S_{I^*-n^{\d/20}}$ in $\H$ satisfying Lemma~\ref{lTw}.
Our strategy will be to show that Property~\ref{pfvu} holds for $u^*$ and the $n^{\d/20}$ variables on the directed path $T(w,u^*)$.  Thus we want to show that for each vertex $v$ on this path, $\chi(v)$ contains no free vertices other that $u^*$.  Lemma~\ref{lgap} will establish that there are no such free vertices outside of the 2-core. So we will begin by arguing that \aas\ each $\chi(v)$ contains no free vertices inside the 2-core; i.e.\ the free vertices on core flippable cycles.

 By Lemma~\ref{lTw}(b) at most $n^{3\delta/2}$  vertices in $\C_2=\C_2(\H)$ are adjacent to $T(w)\setminus \C_2$. We first prove that a.a.s.\ none of these  vertices in $\C_2$ are contained in a core flippable cycle of $\H$, and thus a.a.s.\ no vertex of any core flippable cycle  is in $T(w)$. 

We can choose $\H$ in the following way.  First, choose a random hypergraph $\H_1=\H_r(n,cn)$. Then form $\H_2$ by randomly permuting the vertices of the 2-core of $\H_1$.  That is: let $\s$ be a uniformly random permutation of the vertices of the 2-core.  Replace every hyperedge $(v_1,...,v_r)$ in the 2-core with $(\s(v_1),...,\s(v_r))$, and keep every hyperedge with at least one non-2-core vertex unchanged.  Note that the vertex set of $\C_2(\H_2)$ is equal to the vertex set of $\C_2(\H_1)$; to see this, consider any stripping sequence which when produces the 2-core of $\H_1$ - the same sequence will produce the same set of vertices as the 2-core of $\H_2$. 

We claim that $\H_2$  is distributed like  $\H_r(n,cn)$ and hence is a valid choice for $\H$.
To see this, partition the set of hypergraphs with $cn$ edges into equivalence classes where $\H\sim\H'$ if $\C_2(\H),\C_2(\H')$ are isomorphic and have the same vertex set.  The procedure in the previous paragraph first chooses an equivalence class with probability proportional to its size, and then chooses a uniform member of that class.  Thus it picks a uniform hypergraph with $cn$ hyperedges.

{As we said above, at most $n^{3\delta/2}$ vertices in $\C_2(\H_1)$ are adjacent to $T(w)\setminus \C_2$.} 
By Lemma~\ref{lflip}, the total number of vertices contained in core flippable cycles of $\C_2(\H_1)$ is $O(n^{\d/2}\log^2 n)$. So after taking the random permutation, the probability that any of the vertices adjacent to $T(w)\setminus \C_2$ is contained in a core flippable cycle of $\H_2$ is $O(n^{3\delta/2+\d/2-1}\log^2 n)=o(1)$, as $\delta<1/2$. This confirms that a.a.s.\  $T(w)$ contains no vertex of a core flippable cycle of $\H_r(n,cn)$.

Now consider any vertex $v\in T(w,u^*)$.
Since $T(w,u^*)$ induces a directed path in $\cald$ by Lemma~\ref{lTw}(c), a.a.s.\ there is exactly one directed path from $u^*$ to $v$, and it follows easily that $u^*\in \chi(v)$. Since $T(v)\subseteq T(w)$, a.a.s.\ no vertex of any core flippable cycle is in $T(v)$, as a.a.s.\ none is in $T(w)$.  By Lemma~\ref{lgap}, $u^*$ is the only vertex in $\cup_{i\geq I^*-n^{\d/20}} S_i$
with indegree zero in $\cald$.
Therefore, $u^*$ is the only free variable in $T(v)$ and so $\chi(v)=\{u^*\}$, as we wanted to prove.

Therefore, if any two solutions  in the same cluster differ on $u^*$ then they differ on all of the
$n^{\d/20}+1$ variables on the path from $u^*$ to $w$ by Lemma~\ref{lTw}(c).  Since every cluster contains one solution for each setting of the free variables, this implies that every cluster is not $n^{\d/20}$-connected. This completes the proof for Theorem~\ref{tc1}(c).
\end{proof}

\subsection{The parallel stripping process and SLOW-STRIP}
\lab{sec:parallel}

To prove Lemmas~\ref{lgap} and~\ref{lTw}, we will work in the AP-model. Corollary~\ref{cor:AP} then implies that these Lemmas hold for $\calh_r(n,cn)$.

In this subsection, we state some results on the parallel 2-stripping process including the number of iterations the process takes and the changes of $|S_i|$ in each iteration.

 Note that we only need to consider $c={c_{r}^*}+n^{-\d}$ by the hypotheses of Theorem~\ref{tc1}.

The following theorem, from~\cite[Theorem 3]{gm}, bounds the number of iterations the parallel 2-stripping sequence takes. The original statement was for $\calh_r(n,cn)$ and holds for any $k$-stripping sequence where $(k,r)\neq (2,2)$, but the proof used $AP_r(n,cn)$ and Corollary~\ref{cor:AP}.

\begin{theorem}\lab{mt}
Fix $\dd\ge 3$. For any constant $0<\d<1/2$ and $c=c_{r}^*+n^{-\d}$,  \aas\ the number of iterations the parallel 2-stripping process takes, when it is applied to $AP_{\dd}(n,cn)$, is $\Theta(n^{\d/2}\log n)$.
\end{theorem}

Next, we state some properties of $|S_i|$ in the parallel 2-stripping process, applied to $AP_r(n,cn)$, where $c={c_{r}^*}+n^{-\d}$ and $0<\d<1/2$. These properties follows from~\cite[Lemma 49]{gm}.

\begin{lemma}\lab{lsi} {For every $\eps>0$,} there exist constants $B,Y_1,Y_2,Z_1$, {dependent only on $r, \eps$,} such that \aas\ {$\sum_{i\ge B}|S_i|\le \eps n$, and} for every $ B\leq i< \imax$ with $|S_i|\ge n^{\d}\log^2 n$:
\begin{enumerate}
\item[(a)] if $|S_i|< n^{1-\d}$ then $(1-Y_1n^{-\d/2})|S_i|\leq |S_{i+1}|\leq  (1-Y_2n^{-\d/2})|S_i|$;
\item[(b)] if $|S_i|\geq n^{1-\d}$ then $\left(1-Y_1\sqrt{\frac{|S_i|}{n}}\right)|S_i|\leq |S_{i+1}|\leq  \left(1-Y_2\sqrt{\frac{|S_i|}{n}}\right)|S_i|$;
\item[(c)] $\sum_{j\ge i}|S_j|\le {Z_1}|S_i|n^{\d/2}$;
\item[(d)] the maximum degree of $AP_r(n,cn)$ is at most $\log n$.
\end{enumerate}
\end{lemma}

In each step of SLOW-STRIP  on $AP_r(n,cn)$,  one vertex-copy of the vertex {$u$} at the front of $\Q$ is deleted, together with another $r-1$ vertex-copies that are in the same part (i.e. hyperedge) of the removed vertex-copy. {Since $u$ has only one vertex-copy, it is deleted from $\Q$ by the end of the step.} For each $i$, we define: 

\smallskip

 $t(i)$ is the step in SLOW-STRIP {during which the first vertex in $S_i$ reaches  the front of $\Q$ (it may be deleted in this step if it happens to have degree zero)}.
 
{Roughly speaking,} $t(i)$ is the step of SLOW-STRIP corresponding to the beginning of the $i$-th iteration of the parallel stripping process.

\bigskip

We need to bound on the rate at which heavy vertices become light during SLOW-STRIP.   As the number of light vertices gets small, we will typically remove exactly one light vertex in each step.  So we want the rate at which new light vertices arise to be less than one.   It is well-known that when $c$ is close to $c_r^*$, that rate is close to one;  much of the work in~\cite{gm} was to bound it away from one. The following bound comes from the analysis in~\cite{gm} but is not stated explicitly there:

\begin{lemma}\lab{br} There is a constant $K>0$ such that a.a.s.\  at every step of SLOW-STRIP, the degrees of the remaining vertices are such that the expected number of heavy vertices that become light in that step is at most 
$1-Kn^{-\d/2}$.
\end{lemma}

\proofstart  In each such step, we remove the only remaining copy of the light vertex at the front of $\Q$ and $r-1$ uniformly chosen copies. The proof of Lemma 16 in~\cite{gm} establishes that each time we remove a uniformly chosen copy, the probability that a heavy vertex becomes light (i.e.\ that we remove a copy of a degree two vertex) is at most $1/(r-1)-\Theta(n^{-\d/2})$.
(See Definition 30, line (34) and the discussion preceding line (29) of~\cite{gm} and let $k=2$.)  This yields the lemma.
\proofend

We will briefly highlight the key arguments that lead to this bound.  Near the end of the process, what remains is the 2-core plus a few extra vertices.  So the degree distribution is very close to that of the 2-core.  When $c$ is close to $c_r^*$ the average degree of the 2-core is very close to a particular value $c'_r$ such that: the proportion of vertex-copies from a 2-core of density $c'_r$ which are in vertices of degree two is very close to $1/(r-1)$.    Using an equation which relates  the average degree of the 2-core of $AP_r(n,cn)$ to $c$, it was shown in~\cite[Lemma 8]{gm} that if $c=c^*_r+n^{-\d}$ then the average degree of the 2-core is roughly $c'_r+n^{-\d/2}$. It followed that the proportion of vertex-copies of that 2-core which are in bins of size two is roughly $1/(r-1)-\Theta(n^{-\d/2})$ (see~\cite[Lemma 16]{gm} for a detailed proof).

\subsection{Proof of Lemmas~\ref{lgap} and~\ref{lTw}}\lab{twoLemmas}

We will prove that Lemmas~\ref{lgap} and~\ref{lTw} hold in the AP-model.  Corollary~\ref{cor:AP} then shows that they hold for $X_r(n,cn)$.

\begin{proof}[Proof of Lemma~\ref{lgap}]
Choose $\n=\hf - \d/6$. Run the parallel $2$-stripping process and let $i_1$ denote the first iteration in which  $|S_{i_1}|\leq n^{\n}$.   Our first step will be to show
that \aas\ $I^*> i_1+n^{\d/20}$.

Let $i_2= i_1+n^{2\d/5}$ and $i_3= i_1+2n^{2\d/5}$.
 By Lemma~\ref{lsi}(c),  a.a.s.\ $\sum_{j\geq i_1} |S_{j}|=O(|S_{i_1}|n^{\d/2})=O(n^{\hf+\d/3})=o(n^{1-\d})$ for small $\d$.
Thus, \aas\ at any iteration $i\geq i_1$, the total number of light vertices (vertices with degree less than $2$) is $o(n^{1-\d})$.

Therefore, we can apply Lemma~\ref{lsi}(a) ({recursively}) to obtain that for all $i_1\leq i\leq i_3$:
\be\lab{br2}
|S_{i}|>\left(1-Y_1n^{-\d/2}\right)^{2n^{2\d/5}}|S_{i_1}|=(1-o(1))|S_{i_1}|>\hf n^{\n}.
\ee
This is valid since in each recursion $i_1\le i\le i_3$, we have $|S_i|=\Omega(|S_{i_1}|)\ge n^{\d}\log^2 n$, provided $\d<3/7$, and so the assumption of Lemma~\ref{lsi} is satisfied. Lemma~\ref{lsi}(a) also implies that for all ${i_1\le i\le i_3}$, 
\be\lab{br3}
|S_i|=O(n^{\n}).
\ee

Our proof consists of two steps. First, we show that $I^*\geq i_2$  by proving:

{\em Claim 1:} A.a.s.\ there is a free variable in $\cup_{i_2\leq i\leq i_3}S_i$.

Then we prove that  after level $i_1$, the free variables are all separated by many levels:

{\em Claim 2:} A.a.s. there is no  integer $i> i_1$ such that $\cup_{i'=i}^{i+n^{\d/20}}S_{i'}$ contains two free variables.

These prove the lemma as follows. Claim 1 implies that $I^*\ge i_2>i_1+n^{\d/20}$. So if there was another level $S_i$ containing a free variable with $i\geq I^*-n^{\d/20}$, then this would violate Claim 2. 

{\em Proof of Claim 2.} Recall how SLOW-STRIP is run in the AP-model:   Since $k=2$, every light vertex, i.e. every vertex in $\msq$, has exactly one copy remaining.  At each step, the copy of the vertex at the front of $\msq$ is removed, along with $r-1$ vertex-copies selected uniformly from amongst all remaining copies; we call each of these $r-1$ selections a {\em trial}. A light vertex $v$ becomes a free variable iff its remaining copy is chosen for removal before it reaches the front of $\msq$.  When a light vertex $u$ selects the remaining copy of a light vertex $v$ during one of its $r-1$ trials, we say that $u$ {\em frees} $v$.  

We let $E_S$ denote the event that for all $i\geq i_1$, we have $|S_i|\leq n^{\n}$;  the definition of $i_1$ and Lemma~\ref{lsi}(b) ensure that $E_S$ holds \aas.
Note that when the last member of $S_i$ is removed from $\msq$, then $\msq$ consists all members of $S_{i+1}$.  So $\msq$ can only contain members from two consecutive levels  $S_i,S_{i+1}$. Thus if $E_S$ holds then 
for all $t\geq t(i_1)$, i.e.\ for all steps of SLOW-STRIP that correspond to iterations $i\geq i_1$ of the parallel process, we always have $|\msq|<2n^{\nu}$.

Now consider any $i> i_1$.  If there are two free variables in $\cup_{i'=i}^{i+n^{\d/20}}S_{i'}$ then there are  two different trials in which a vertex is freed by vertices in $\cup_{j=i-1}^{i+n^{\d/20}}S_{j}$.  If $E_S$ holds then there are at most $(n^{\d/20}+2)n^{\nu}(r-1)$ such trials and during each such trial, there are at most $2n^{\nu}$ vertex-copies in $\msq$, i.e.\ vertex-copies whose choice would free a vertex.  By Theorem~\ref{tkim}(b), we can assume that there are always at least $r\b(c) n$ remaining vertex-copies. So in each trial, the probability of freeing a vertex is at most $2n^{\nu}/r\b(c) n$.   Putting this together, the probability that $E_S$ holds and there are two free variables in $\cup_{i'=i}^{i+n^{\d/20}}S_{i'}$ is at most:
\[ {(n^{\d/20}+2)n^{\nu}(r-1)\choose 2}\left(\frac{2n^{\nu}}{r\b(c) n}\right)^2<n^{-.55\d},\]
since $\n=\hf-\d/6$.
Since $E_S$ holds \aas, it follows that \aas\ there is no such $i$ in the range $i_1\leq i\leq \imax$, where $\imax$ is the last iteration of the parallel stripping process, which is a.a.s.\ less than $n^{0.51\d}$ by Theorem~\ref{mt}. This proves the claim.

{\em Proof of Claim 1}.
Let $E_T$ denote the event that for all $i_2\leq i\leq i_3$ we have $|S_i|\geq \hf n^{\n}$. $E_T$ holds \aas\ by~(\ref{br2}).

If at least one vertex is freed during the removal of the vertices in $\cup_{i_2\leq i\leq i_3-1}S_i$, then there is a free variable in $\cup_{i_2\leq i\leq i_3}S_i$.  If $E_T$ holds, then the total number of vertices removed is at least
$(i_3-i_2)\hf n^{\n}>n^{\n+2\d/5}$.  

Consider the removal of the first $\hf|S_i|$ vertices of $S_i$.  If no vertices are freed during these removals, i.e.\ if each time we remove the first member of $\msq$, we don't select any other member of $\msq$ for deletion, then the size of $\msq$ remains at least $\hf|S_i|>\inv{4}n^{\n}$ if $E_T$ holds.  If there are no free variables in $S_i$, then every member of $\msq$ has degree 1 (not 0) and so each deletion of a member of $\msq$ results in  $r-1$ trials. At each such trial, the total number of vertex-copies remaining is at most $rcn$. It follows that the probability that $E_T$ holds and no vertex is freed during the removal of the first $\hf|S_i|$ vertices over all $i_2\leq i\leq i_3$ is at most
\[\left(1-\frac{n^{\n}/4}{rcn}\right)^{\hf (r-1)n^{\n+2\d/5}}=e^{-\Theta(n^{2\d/5+2\n-1})}=o(1),\]
since $\n=\hf-\d/6$.  Because $E_T$ holds \aas, this proves Claim 1 and so completes the proof of the lemma.
\end{proof}

\begin{proof}[Proof of Lemma~\ref{lTw}]
 As in the previous proof, we let  $\n=\hf - \frac{\d}{6}$. Run  SLOW-STRIP {on $AP_r(n,cn)$}  and let $i_1$ denote the first iteration in the parallel stripping process such that  $|S_{i_1}|\leq n^{\n}$; set $t_1=t(i_1)$. Recall that $\calc_2$ is the 2-core of $AP_r(n,cn)$.
We will prove: 

\begin{claim}\lab{claim:Tw}
  A.a.s.\ for every $w\in \cup_{i\geq i_1}S_i$, {$T(w)\setminus \calc_2$} induces a directed tree rooted at $w$, where a {\em directed rooted tree} means a tree where every edge is directed towards the root.
\end{claim}

Note that every vertex in any $S_i$ must lie in a hyperedge with a vertex in $S_{i-1}$ that was deleted during iteration $i-1$ of the parallel stripping process.  Thus each vertex in $S_i$ points to a vertex in $S_{i-1}$ in $\cald(\Psi)$. It follows that there is a path in $\cald(\Psi)$ from $u^*$  to some vertex $w\in S_{I^*-n^{\d/20}}$ where that path contains a vertex in each  $S_i$, $I^*-n^{\d/20}\le i\le I^*$.
The claim will imply parts (a,c) for this choice of $w$.

We start by bounding the size of  each $T(w)\setminus\calc_2$.

Fix some $w\in S_i$, with $i\geq i_1$. We maintain a set $\calt(w)$ as follows.

Initially $\calt(w):=\{w\}$. Whenever we delete a  vertex $v\in\msq$ such that $v\in\calt(w)$:
(i) each neighbour of $v$ that has degree at most $2$, and hence is in or will enter $\msq$, is placed in $\calt(w)$; (ii) each neighbour of $v$ that has degree greater than $2$ is coloured Red.
Every time a Red vertex enters $\msq$, it is placed in $\calt(w)$.  Thus, when we finish SLOW-STRIP, $\calt(w)=T(w)\setminus\calc_2$.

We will analyze $\calt(w)$ using a branching process.  {When a vertex $v\in\calt(w)$ is deleted by SLOW-STRIP, we say that we are {\em processing } $v$.  If a vertex $u$ is added to $\calt(w)$ while $v$ is being processed then we consider $u$ to be an {\em offspring} of $v$.
If a vertex $u$ is added to $\calt(w)$ during the deletion of a vertex not in $\calt(w)$ (and so $u$ must be Red), then we consider $u$ to be an {\em offspring} of the most recently processed member of $\calt(w)$.
Note that the offspring of $v$ are not neccessarily adjacent to $v$ in $\cald$. }

In other words: we say that $u$ is an {\em offspring} of a vertex $v\in \calt(w)$, if $u$ entered $\calt(w)$ between the iterations of SLOW-STRIP ranging from the time we remove $v$ up until just before the next iteration where we remove a member of $\calt(w)$. Our definition of an offspring may look a little unnatural. This is because we want to include in $\calt(w)$ those vertices that become light because of the removal of some vertices not in $\calt(w)$, but have already been found to be adjacent to some vertex in $\calt(w)$. Note that these vertices are in $T(w)\setminus \C_2$.

There are two scenarios
under which $u$ can become an offspring of $v$: (i) at the time we delete $v$, $u$ is a neighbour of $v$ and $u$ has degree at most 2; (ii) at the step after $v$ is deleted, $u$ is Red  and  $u$ enters $\msq$ before the next member of $\calt(w)$ is deleted.  For case (ii) to occur, $u$ must be the neighbour of another vertex $v'\notin\calt(w)$ that is removed from $\msq$, and the degree of $u$ must drop below 2 when $v'$ is removed.

It will be convenient to consider $\calt'(w)\subseteq\calt(w)$, which differs from $\calt(w)$ in that only the first $n^{2\d}$ vertices to be coloured Red can enter $\calt'(w)$.  We will show that, in fact, a.a.s.\ $\calt'(w)=\calt(w)$.

When removing  $v\in\calt'(w)$,
the expected number of offspring created under scenario (i) is {at most} $1-Kn^{-\d/2}$ for some constant $K>0$ by Lemma~\ref{br}, and this holds for all $t\geq t(i)\geq t(i_1)$.

The number of iterations until the next member of $\calt'(w)$ is removed is at most $O(|S_i|\times n^{\d/2})=O(n^{\n+\d/2})$, by Lemma~\ref{lsi}(c). If $u$ is an offspring of $v$ created under scenario (ii), then $u$ must be one of the first $n^{2\d}$ Red vertices. Furthermore during one of those iterations,  exactly two vertex-copies of $u$ remain and one of them is selected.  The total number of such vertex-copies over all choices of $u$ is  at most $2n^{2\d}$, and since there are a linear number of vertex-copies to choose from, the expected number of offspring of $v$ created under scenario (ii) is at most
\[O(n^{\n+\d/2})\times O(n^{2\d}/n)=O(n^{\n+5\d/2-1})=o(n^{-\d/2}),\]
for sufficiently small $\d$.  

Therefore, the total expected number of offspring of $v$, in $\calt'(w)$, is  at most $1-Kn^{-\d/2}+o(n^{-\d/2})\le 1-z$ for  $z=(K/2)n^{-\d/2}$.  So $\calt'(w)$
grows like a Galton-Watson branching process with branching parameter at most $1-z$.  The probability that such a branching process has size at least $x$ drops quickly as $x$ exceeds $\Theta(z^{-2})$ (see, eg.~\cite{bbckw}), and in particular, $\Pr(|\calt'(w)|>n^{3\d/2}/(r-1))=o(1/n)$.
So \aas\ $|\calt'(w)|\leq n^{3\d/2}/(r-1)$ for every $w$.

Note that at most $r-1$ Red vertices are formed each time a member of $\calt'(w)$ is removed.
So \aas\ the number of Red vertices is at most $(r-1) {|\calt'(w)|}\le n^{3\d/2}<n^{2\d}$ and so $\calt(w)=\calt'(w)$ for all $w$. Therefore \aas\
\[|T(w)\setminus\calc_2|=|\calt(w)| \leq n^{3\d/2}{/(r-1)}\ \mbox{ for every } w\in \cup_{i\geq i_1}S_i.\]
{Since each vertex in $T(w)\setminus \C_2$ can be adjacent to at most $r-1$ vertices in $\C_2$, it follows that a.a.s.\ $|T(w)|\le n^{3\d/2}\ \mbox{for every}\ w\in \cup_{i\geq i_1}S_i$.
This implies that the choice of $w$ below Claim~\ref{claim:Tw} satisfies Lemma~\ref{lTw}(b).
}

Now we prove Claim~\ref{claim:Tw}, i.e.\ we show that \aas\ each $T(w)\setminus\calc_2$ induces {a directed tree rooted at $w$} in $\cald$.

{\bf Observation:}  If $T(w)\setminus\calc_2$ does not induce a directed tree rooted at $w$ in $\cald$, then there must have been an iteration where some $v\in\calt(w)$ is deleted, and one of the $r-1$ neighbours of $v$ {in the remaining hypergraph} was either Red or in $\calt(w)$.

Again, it will be convenient to consider $\calt'(w)$ rather than $\calt(w)$.

When we delete $v$, we choose $r-1$ uniform vertex-copies as its neighbours. Each vertex of $\calt'(w)$ has entered $\msq$ and so has at most one copy remaining. So the probability that we choose  a copy of a vertex in $\calt'(w)$  is at most $|\calt'(w)|/\Theta(n)=O(n^{3\d/2 - 1})$.
So the probability that this happens during the deletion of at least one member of $\calt'(w)$ is $O(n^{3\d/2})\times O(n^{3\d/2 - 1})=O(n^{3\d - 1})$.

To bound the probability of choosing a copy of a Red vertex, note that the total number of copies of Red vertices is a.a.s.\ $O(n^{2\delta}\log n)$, because $\calt'(w)$ allows only at most $n^{2\d}$ vertices to be coloured Red, and a.a.s.\ the degree of each vertex is less than $\log n$. Since the $r-1$ vertex-copies (i.e.\ the neighbours of $v$) are uniformly chosen from the remaining $\Theta(n)$ vertex-copies, the probability that one of them is a copy of a Red vertex is $O(n^{2\d-1}\log n)$. Therefore, the probability that during the deletion of vertices in $\calt'(w)$, there is such $v$ such that one of its $r-1$ neighbours was Red is $O(n^{3\d/2})\times O(n^{2\d - 1}\log n)=O(n^{5\d/2 - 1}\log n)$.

So the probability that, upon deleting some $v\in\calt'(w)$, one of the $r-1$ neighbours is either Red or in $\calt'(w)$ is 
$O(n^{3\d - 1})+O(n^{5\d/2 - 1}\log n) =O(n^{3\d - 1})$.
 Multiplying by the $O(n^{\n+\d/2})$ choices for $w\in\cup_{i\geq i_1}S_i$ (by Lemma~\ref{lsi}(c))) we get $O(n^{\n+7\d/2-1})=o(1)$
for $\d$ sufficiently small.  
This proves that  \aas\  there is no iteration where we delete some $v\in\calt'(w)$ and one of the $r-1$ neighbours of $v$ is either Red or in $\calt'(w)$.  We proved above that \aas\ $\calt(w)=\calt'(w)$ for all $w$, and so the same is true for $\calt(w)$.  So our Observation proves that \aas\    $T(w)\setminus\calc_2$ induces {a directed tree rooted at $w$} in $\cald$ for every $w\in\cup_{i\geq i_1}S_i$, which proves (a,b,c) as described above.

\end{proof}

\section{Concluding remarks}

We have examined the solution clusters of $X_r(n,cn)$ for $c=c_r^*+o(1)$.  We showed that for small constant $\d>0$: when $c=c_{r}^*-n^{-\d}$, \aas\ all solutions are $n^{O(\d)}$-connected; whereas when $c=c_{r}^*+n^{-\d}$, \aas\ the connectivity parameter of each cluster is $n^{\Theta(\d)}$, and different clusters are $\Omega(n^{1-r\d})$-separated. This indicates a rather smooth cluster transition near the clustering threshold.

It is possible that the clusters are even more separated than we have shown. We would like to know whether the clusters are a.a.s.\ pairwise $\Omega(n)$-separated; or all solutions of $X_r(n,cn)$ are a.a.s.\ $o(n)$-connected.

When $c=c_{r}^*+n^{-\d}$ we have shown that a.a.s.\ each cluster contains two solutions that are $n^{z\d}$-separated for some constant $z>0$. We conjecture that this holds as well when $c=c_{r}^*-n^{-\d}$ (see Conjecture~\ref{conj}). However, our proof technique for $c=c_{r}^*+n^{-\d}$ does not apply to $c=c_{r}^*-n^{-\d}$.

More importantly, we would like to see what happens when $\d>\hf$ and so we do not know that a 2-core arises $\whp$.  In that setting, consider creating a sequence of random hypergraphs by starting with $n$ vertices and adding uniform random hyperedges one at a time. We would like to find out whether clusters arise as soon as the first non-empty 2-core appears in the underlying hypergraph. Is there always some $f(n)=o(g(n))$ such that if there is a 2-core then, under the cluster definitions from Section~\ref{ssc}, any two solutions in the same cluster are $f(n)$-connected, while any two solutions in different clusters are $g(n)$-separated?  Or do we need to change the definition of clusters?  Or perhaps our notion of clustering falls apart at the very moment when the 2-core is formed.

Instances of $r$-XORSAT can be solved in polynomial time, using global algorithms such as Gaussian elimination.  However, when $c$ is above the clustering threshold, random $r$-XORSAT seems to be very difficult for generic CSP solvers and local algorithms such as WalkSat\cite{young}.  It is natural to wonder whether such difficulties arise precisely {at the step when} the 2-core appears.  Answering this question, and understanding exactly what these difficulties are, could provide insights into how it is that  clustering creates algorithmic difficulties for other random CSP's.  Resolving the issues discussed in the previous paragraph would be very helpful for this question.


\begin{thebibliography}{99}



\bibitem{aco}
D. Achlioptas and A. Coja-Oghlan.
{\em Algorithmic barriers from phase transitions.}
 In 49th Annual IEEE Symposium on Foundations of Computer Science,
               FOCS 2008, October 25-28, 2008, Philadelphia, PA, USA, pages 793--802. IEEE Computer Society, 2008.


\bibitem{amxor} D. Achlioptas and M. Molloy, The solution space geometry of random linear equations, {\em Random Structures \& Algorithms} 46(2): 197--231, (2015). 	arXiv:1107.5550 .

\bibitem{art}
D. Achlioptas and F. Ricci-Tersenghi,
{Random formulas have frozen variables.}
{\em SIAM J. Comput.},
{39} (1),
260--280 (2009).




\bibitem{bmz} A. Braunstein, M. Mezard and R. Zecchina, {Survey propagation: an algorithm for satisfiability}, {\em Random Structures \& Algorithms} {27}: {201--226}, (2005).




\bibitem{bb3} B. Bollob\'{a}s, {The evolution of random graphs}, {\em Transactions of the AMS}, {286}: {257--274}, (1984).

\bibitem{bbckw}  B. Bollob\'{a}s, C. Borgs, J. Chayes, J. Kim and D. Wilson, {The scaling window of the 2-SAT transition}, {\em Random Structures \& Algorithms} {18}: {201--256}, (2001).

\bibitem{CW} Julie Cain and Nicholas Wormald, Encores on cores,
{\em Electron. J. Combin.}, 13(1), Research Paper 81, 13 pp, (2006).


\bibitem{nc} N. Calkin, {Dependent sets of constant weight binary vectors}, {\em Combinatorics, Probability and Computing}, {6}: {263--271}, (1997).



\bibitem{Chvatal} V. Chv\'{a}tal,  Almost all graphs with $1.44 n$ edges are 3-colorable, {\em Random Structures \& Algorithms}, 2(1):{ 11--28}, (1991).

\bibitem{cdmm} S. Cocco, O. Dubois, J. Mandler and R. Monasson, {Rigorous decimation-based construction of ground pure states for spin glass models on random lattices.}
{\em Phys. Rev. Lett.} {90}, 047205, (2003).


\bibitem{coalg}  A. Coja-Oghlan, {A better algorithm for random {k-SAT}}, {\em SIAM Journal on Computing} {39}: {2823--2864}, (2010).

\bibitem{coind} A. Coja-Oghlan and C. Efthymiou, {On independent sets in random graphs}, {\em Proc. 22nd SODA}, {136--144}, (2011).

\bibitem{cop} A. Coja-Oghlan, K. Panagiotou, {Catching the $k$-NAESAT threshold}, {\em Proc. 44th STOC}, {899--908},  (2012).

\bibitem{cop2} A. Coja-Oghlan, K. Panagiotou, {Going after the $k$-SAT threshold}, {\em Proc. 45th STOC}, {705--714},  (2013).

\bibitem{cov} A. Coja-Oghlan, D. Vilenchik, {Chasing the $k$-colorability threshold}, {\em Proc. FOCS}, 380--389, (2013).

\bibitem{cz} A. Coja-Oghlan and L. Zdeborova, {The condensation transition in random hypergraph 2-coloring}, {\em Proc. 23rd SODA}, {241--250}, (2012).

\bibitem{Coja} A. Coja-Oghlan, The asymptotic $k$-sat threshold, {\em Proc. STOC}, 804--813, ACM, (2014).


\bibitem{cdxor} N. Creignou and H. Daud\'e,  {Satisfiability threshold for random XOR-CNF formulas}, {\em Discrete Appl. Math.}, {96-97}: {41--53}, (1999).
\bibitem{daud}
H. Daud{\'e}, M. M{\'e}zard, T. Mora, and R. Zecchina,
{Pairs of SAT assignments and clustering in random boolean formulae},
{\em Theoretical Computer Science},
{393}(1-3):
{260--279}, (2008).


\bibitem{cuc}
M. Dietzfelbinger, A. Goerdt, M. Mitzenmacher, A. Montanari, R. Pagh and M. Rink,
{Tight thresholds for cuckoo hashing via XORSAT}, {\em Proceedings of  Automata, Languages and Programming, 37th International
               Colloquium, ICALP}, {213--225},  2010.






\bibitem{dub}
O. Dubois and J. Mandler,
{The 3-XORSAT threshold}, {\em Comptes Rendus Mathematique}, 335(11): 963--966, (2002).




\bibitem{dg} D. Gamarnik and M. Sudan, {Limits of local algorithms over sparse random graphs}, {\em Proceedings of the 5th conference on Innovations in theoretical computer science} 369--376, ACM, (2014). 


\bibitem{DSS} J. Ding, A. Sly, and N. Sun,  Proof of the satisfiability conjecture for large k, arXiv:1411.0650, (2014).

\bibitem{fr} D. Fernholz and V. Ramachandran, The giant $k$-core of a random graph with a specified degree sequence.  Unpublished manuscript (2004).

\bibitem{gm} P. Gao and M. Molloy, { The stripping process can be slow: part I}, arXiv: 1501.02695.

\bibitem{gm2} P. Gao, { The stripping process can be slow: part II}, arXiv: 1505.02804.


\bibitem{young}
M. Guidetti and A.P. Young, {Complexity of several constraint-satisfaction problems using the heuristic classical algorithm WalkSAT}, {\em Phys. Rev. E}, {84} (1), 011102, 2011.


\bibitem{ikkm}
M. Ibrahimi, Y. Kanoria, M. Kraning, and A. Montanari,
{The set of solutions of random xorsat formulae},
{\em Proc. SODA}, {760--779}, SIAM, 2012.

\bibitem{jlkp}
S. Janson, T. {\L}uczak, D. Knuth, and B. Pittel, {The birth of the giant component}, {\em Random Structures \& Algorithms} {3}: {233--358}, (1993).


\bibitem{jhk} J.H.Kim,  {Poisson cloning model for random graphs},  {\em International Congress of Mathematicians} Vol. III, 873--897, Eur. Math. Soc., Z\"{u}rich, 2006. 

\bibitem{vk1} V. Kolchin, {Random graphs and systems of linear equations in finite fields}, {\em Random Structures \& Algorithms}, {5}: {135--146}, (1994).
\bibitem{vk2} V. Kolchin and V. Khokhlov, {A threshold effect for systems of random equations of a special form}, {\em Discrete Mathematics and Applications}, {5}:  {425--436}, (1995).


\bibitem{kmrsz} F. Krzakala, A. Montanari, F. Ricci-Tersenghi, G. Semerjian and L. Zdeborova, {Gibbs States and the Set of Solutions of Random Constraint Satisfaction Problems}, {\em Proc. Natl. Acad. Sci.}, 104(25): 10318-10323, (2007).




\bibitem{tlcomp}  T. {\L}uczak,  {Component behaviour near the critical point of the random graph process},
{\em Random Structure \& Algorithms}, {1}: {287--310}, (1990).


\bibitem{lpw}  T. {\L}uczak, B. Pittel and J. Weirman, {The structure of a random graph at the point of the phase transition}, {\em Trans. Am. Math. Soc.} {341}:  {721--748}, (1994).



\bibitem{mmbook}

M. M{\'e}zard and A. Montanari.
\newblock {\em Information, Physics, and Computation}.
\newblock Oxford University Press, Inc., New York, NY, USA, 2009.

\bibitem{mmz}
M. M{\'e}zard, T. Mora,  and R. Zecchina,
{Clustering of solutions in the random satisfiability problem},
{\em Phys. Rev. Lett.},
{94}(19),
197205 (2005).

\bibitem{sp}
M. M{\'e}zard, G. Parisi, and R. Zecchina,
{Analytic and algorithmic solution of random satisfiability problems},
Science,
{297}:
{812--815}, (2002).


\bibitem{mez}
M. M{\'e}zard, F. Ricci-Tersenghi, and R. Zecchina, {Two solutions to diluted $p$-spin models and XORSAT
problems}, {\em J. Stat. Phys.} {111}: {505--533}, (2003).


\bibitem{mz}
M. Mezard, R. Zecchina, {The random K-satisfiability problem: from an analytic solution to an efficient algorithm}, {\em Phys. Rev. E} {66}, (2002).

\bibitem{mmcore} M. Molloy, {Cores in random hypergraphs and boolean formulas},
{\em Random Structures \& Algorithms}, {27}: {124--135}, (2005).

\bibitem{mmfreeze} M. Molloy, {The freezing threshold for $k$-colourings of a random graph},
{\em Proc. STOC}, 921--930, (2012).





\bibitem{mres} M. Molloy and R. Restrepo, {Frozen variables in random boolean constraint satisfaction problems}, {\em Proc. SODA}, 1306--1318, (2013).



\bibitem{mrt}
A. Montanari, R. Restrepo and P. Tetali, {Reconstruction and clustering in random constraint satisfaction problems}, {\em SIAM J. Disc. Math.}, {25}: {771--808}, (2011).


\bibitem{mpwz} R. Mulet, A. Pagani,  M. Weigt and R. Zecchina,  {Coloring random graphs}, {\em Phys. Rev. Lett.} {89}, 268701, (2002).


\bibitem{ps} B. Pittel and G. Sorkin, {The satisfiability threshold for $k$-XORSAT}, 	arXiv:1212.1905 (2012).



\bibitem{psw} B. Pittel, J. Spencer and N. Wormald,
{Sudden emergence of a giant $k$-core in a random graph},
{\em J. Comb. Th. B} {67}: {111--151}, (1996).



\bibitem{gs} G. Semerjian, {\em On the freezing of variables in random constraint satisfaction problems.} J.Stat.Physics {\bf 130} (2008), 251~-~293.

\bibitem{zk} L. Zdeborov\'a and F. Krzakala, {Phase transitions in the colouring of random graphs},
{\em Phys. Rev. E} 76, 031131, (2007).

\end{thebibliography}
\end{document}